\documentclass[fleqn,usenatbib,a4paper]{mnras}
\usepackage{newtxtext,newtxmath}
\usepackage[T1]{fontenc}
\usepackage{ae,aecompl}
\usepackage{color}

\usepackage{graphicx}	
\usepackage{amsmath}	
\usepackage{amssymb}	

\usepackage{subcaption}
\title[Anisotropic emission of neutrino and GW signals from rapidly rotating CCSNe]{Anisotropic  emission of neutrino and gravitational-wave signals from rapidly rotating core-collapse supernovae}

\author[T. Takiwaki and K. Kotake]{
Tomoya Takiwaki$^{1}$\thanks{E-mail: takiwaki.tomoya@nao.ac.jp} and
Kei Kotake$^{2}$\thanks{E-mail: kkotake@fukuoka-u.ac.jp}\\
$^1$ Division of Theoretical Astronomy, National Astronomical Observatory of 
 Japan, 2-21-1, Osawa, Mitaka, Tokyo, 181-8588, Japan\\
$^2$ Department of Applied Physics, Fukuoka University, Jonan, Nanakuma, Fukuoka 814-0180, Japan}

\date{\today}

\pubyear{2017}

\begin{document}
\label{firstpage}
\pagerange{\pageref{firstpage}--\pageref{lastpage}}
\maketitle

  \begin{abstract}
 We present analysis on neutrino and GW signals based 
on three-dimensional (3D) core-collapse supernova simulations 
of a rapidly rotating 27 $M_{\odot}$ star. 
We find a new neutrino signature that is produced by a 
 lighthouse effect where the spinning of strong neutrino emission
 regions around the rotational axis leads to quasi-periodic 
modulation in the neutrino signal. 
 Depending on the observer's viewing angle,
 the time modulation will be clearly detectable in IceCube and 
the future Hyper-Kamiokande. The GW emission is also anisotropic 
where the GW signal is emitted, as previously identified, 
most strongly toward the equator at rotating core-collapse 
and bounce, and the non-axisymmetric instabilities 
in the postbounce phase lead to stronger GW emission 
toward the spin axis. We show that these GW signals can 
be a target of LIGO-class detectors for a Galactic event.
  The origin of the postbounce GW emission naturally explains 
why the peak GW frequency is about twice of the neutrino 
modulation frequency. We point out that the simultaneous 
detection of the rotation-induced neutrino and GW signatures 
could provide a smoking-gun signature of a rapidly rotating 
proto-neutron star at the birth.
  \end{abstract}




\section{Introduction}\label{sec:introduction}
Detection of neutrinos and gravitational waves (GWs) from core-collapse 
supernovae (CCSNe) has been long
expected to provide
 crucial information for understanding the explosion mechanism 
 (e.g., \citet{mirizzi16,Kotake13}
 for a review). Currently multiple neutrino detectors are in operation,
 where the best suited ones for detecting CCSN neutrinos are
 Super-Kamiokande and IceCube 
(e.g., \citet{scholberg12}). Ever since SN1987A, 
significant progress has been 
 also made in GW detectors. The high sensitivity led to the 
Nobel-prize-awarded detection by the LIGO 
collaboration \citep{gw2016} from the black hole merger event.
Virgo also reported the first joint 
GW detection \citep{joint}. KAGRA will start the operation in the coming years
 \citep{Aso13}. 
Targeted by these multi-messenger observations,
 the importance of neutrino and GW predictions from CCSNe is ever
 increasing.

From the CCSN theory and simulations, it is almost certain that
 multi-dimensional (multi-D) hydrodynamics instabilities including
 neutrino-driven convection and the Standing-Accretion-Shock-Instability 
(SASI)
 play a crucial role in facilitating the 
neutrino mechanism of CCSNe (\citet{bethe}, see \citet{burrows13,janka16} for review). 
In fact, a number of self-consistent 
 models in two or three spatial dimensions (2D, 3D)
 now report revival of the stalled bounce shock into explosion
 by the multi-D neutrino mechanism (see, e.g., 
\citet{bernhard17,bollig17,Roberts16,melson15b,lentz15,Nakamura15}
for collective references therein).

In order to get a more robust and sufficiently energetic explosion 
to meet with observations, some 
physical ingredients may be still missing (e.g., \citet{janka16}).
 Possible candidates to enhance the chance of explodability 
include general relativity (e.g., \citet{BMuller12b,KurodaT12,Roberts16}), 
inhomogenities in the progenitor's burning shells 
(e.g., \citet{Couch15,bernhard15i}),
rotation \citep{Marek09,Suwa10,takiwaki16,summa17} and magnetic fields 
(e.g., \citet{moesta15,jerome15,tomek16,martin17}).

In our previous work, we reported a new 
type of rotation-assisted, neutrino-driven explosion of a $27 M_{\odot}$ star 
\citep{takiwaki16}, where the growth of non-axisymmetric instabilities 
due to the so-called low-T/|W| instability is the key to foster the onset of the 
explosion.
More recently, \citet{summa17}
 found yet another type of rotation-assisted, neutrino-driven explosion of 
 a 15 $M_{\odot}$ star where the powerful spiral SASI motions play a crucial 
role in driving the explosion. Note that the GW signals from rapidly
 rotating core-collapse and bounce 
have been extensively studied so far (e.g., \citet{Dimmelmeier08,Ott09,Scheidegger10,KurodaT14}). The correlation of the neutrino and GW signals
from rapidly rotating models have been studied in 3D models \citep{Ott12a} with 
octant symmetry focusing on the bounce signature
 (limited to $\sim 30$ ms postbounce) and in \citet{Yokozawa} based on their 2D models.
However, the postbounce GW and neutrino signals have not yet been studied 
 in the context of self-consistent, 3D neutrino-driven models 
aided by rotation.

In this Letter, we present analysis of neutrino and GW signals from 
the rotation-aided, neutrino-driven explosion of a $27 M_{\odot}$ star 
in \citet{takiwaki16}.
 We find a new neutrino signature that is produced by a 
 {\it neutrino} lighthouse effect.
 We discuss the detectability in IceCube and the future 
Hyper-Kamiokande. By combining with the postbounce GW signatures, 
 we find that the peak GW frequency is 
physically linked to the peak neutrino modulation frequency. 
We point out that these signatures,
 if simultaneously detected, could provide evidence 
 of rapid rotation of the forming PNS.

\section{Numerical Methods}\label{model}
From our 3D models computed in \citet{takiwaki16}, 
we take the rotation-assisted explosion model of  
$27.0$ $M_{\odot}$ star ("s27.0-R2.0-3D") for the analysis 
in this work. For the model, the constant angular frequency 
of $\Omega_0 = 2$ rad/s is initially imposed to the iron core 
of a non-rotating progenitor of \citet{woosley02} 
with a cut-off ($\propto r^{-2}$) outside. 
We employ
the isotropic diffusion source approximation (IDSA) scheme 
\citep{idsa} for spectral neutrino transport 
of electron-($\nu_e$) and anti-electron-($\bar{\nu}_e$) 
neutrinos and a leakage scheme for heavy-lepton neutrinos ($\nu_x$) 
(see \citet{takiwaki16} for more details).
Note that our 3D run failed to explode in the absence of rotation 
\citep{takiwaki16}, which is in line with \citet{Hanke13} 
who performed 3D full-scale simulations using the same progenitor but 
with more elaborate neutrino transport scheme.

We consider two water \v{C}erenkov neutrino detectors, 
 IceCube \citep{ICECUBE2011,ICECUBE2012} and the future 
Hyper-Kamiokande \citep{HyperK2011,HyperK2016}.
In the detectors, the main detection channel is of anti-electron
 neutrino ($\bar{\nu}_e$) with inverse-beta decay
 (IBD). The observed event rate at Hyper-Kamiokande
 is calculated as follows,
\begin{equation}
 R_{\rm HK} = N_{\rm p}
  \int_{E_{\rm th}} {\rm d} E_e
   \frac{{\rm d} F_{\bar{\nu}_e}}{{\rm d} E_\nu}
   \sigma\left(E_\nu\right)
   \frac{{\rm d} E_\nu}{{\rm d} E_e},
\end{equation}
where $N_{\rm p}=2.95\times 10^{34}$ is the number of protons for
 two tanks of Hyper-Kamiokande whose fiducial volume is designed as 
440 kton,
$E_{\rm th}=7~{\rm MeV}$ is the threshold energy,
$\sigma$ is the cross section of the IBD \citep{fukugita03}.
The neutrino number flux of $\bar{\nu}_e$ ($F_{\bar{\nu}_e}$) at a 
source distance of $D$ is 
estimated as $F_{\bar{\nu}_e} = \frac{\mathcal{L}_{\Omega}}{4\pi D^2}$, where $\mathcal{L}_{\Omega}$ 
denotes an viewing-angle dependent neutrino (number) luminosity (see Appendix A of \cite{tamborra14a}).
In this work, $D$ is set as $10$ kpc unless otherwise stated.  
Following \citet{keil03}, the neutrino spectrum is 
assumed to take a Fermi-Dirac distribution with vanishing 
chemical potential (see their Eq. (5)).

While individual neutrino events cannot be reconstructed,
 a statistically significantly "glow" is predicted to be observed
 during the passage 
 of the CCSN neutrino signals at IceCube (e.g., \citep{mirizzi16}).
We estimate the event rate at IceCube following Equation (2) in 
\citet{lund10}. 

Extraction of GWs from our simulations is done by
 using the conventional quadrupole stress formula 
\citep{Misner73,EMuller97}. For numerical convenience, we employ 
the first-moment-of-momentum-divergence (FMD) formalism proposed 
by \citet{finn90} (see also \citet{Murphy09,BMuller13}, and  \citet{tk11} for
 justification.)

\section{Results}\label{sec:results}

\begin{figure}
\begin{center}
\includegraphics[width=0.99\linewidth]{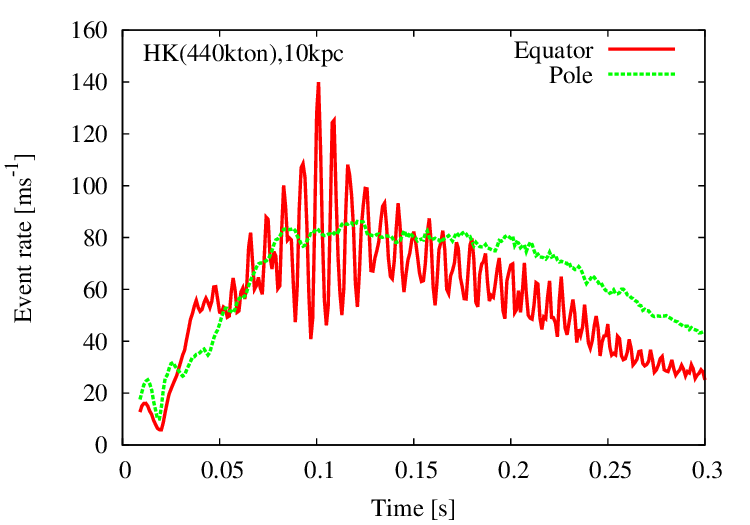}
\includegraphics[width=0.99\linewidth]{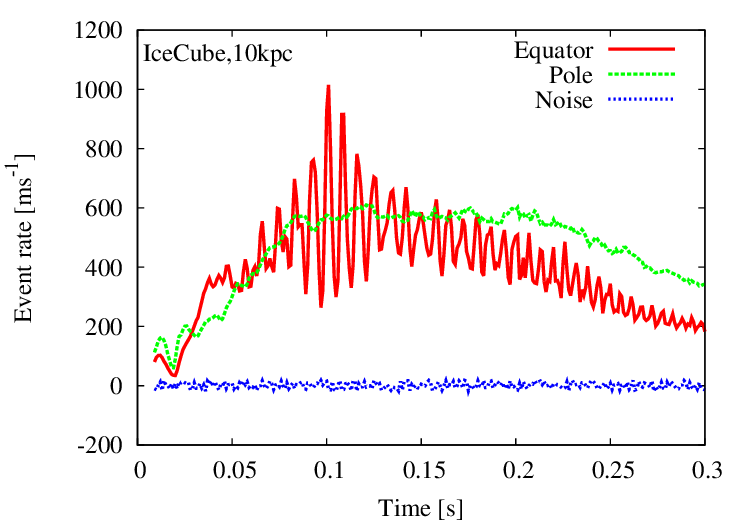}
\end{center}
 \caption{Hyper-Kamiokande (HK, {\it upper}) and IceCube ({\it lower})
 detection rates of $\bar{\nu}_e$ at 10 kpc for our rapidly rotating
 $27 M_{\odot}$ model as a function of time after bounce. 
The red and green line corresponds to the 
 event rates (per 1 ms bin) for an observer along the equator 
 and the pole, respectively.
In both of the detectors, a quasi-periodic modulation (red lines) 
is clearly seen for the observer along the equator.
 For IceCube, the background fluctuations are shown by blue line.}\label{f1}
\end{figure}

 Figure \ref{f1} shows the expected neutrino signals for
our rapidly rotating $27 M_{\odot}$ star at Hyper-Kamiokande 
(top panel) and IceCube (lower panel), respectively.
In both of the detectors, a quasi-periodic modulation (red lines) 
is clearly seen for an observer along the equator, which cannot
 be seen for an observer along the rotational axis (denoted 
as "Pole" in the panel, green lines).
  The modulation amplitude at Hyper-Kamiokande 
(top panel, red line) is significantly higher than the root-N 
Poisson errors, $\sqrt{N}\sim \sqrt{R_{\rm HK}\times 1{\rm ms}}\sim 9$.
 For IceCube, much bigger modulation amplitudes are obtained 
due to its large volume. This is well above the IceCube 
background noise (blue line) 
of $\sim\sqrt{R_{\rm bkgd} \times 1{\rm ms}} 
\sim 38$ for the 1ms bin \citep{tamborra13}.
 For a source at 10 kpc, IceCube plays a crucial role 
 for detecting the signal modulation. But, for the 
 more distant source, Hyper-Kamiokande can be 
 superior because it is essentially background free
 (see \citet{scholberg12} for detail).
 
 \begin{figure}
\begin{center}
\includegraphics[width=0.9\linewidth]{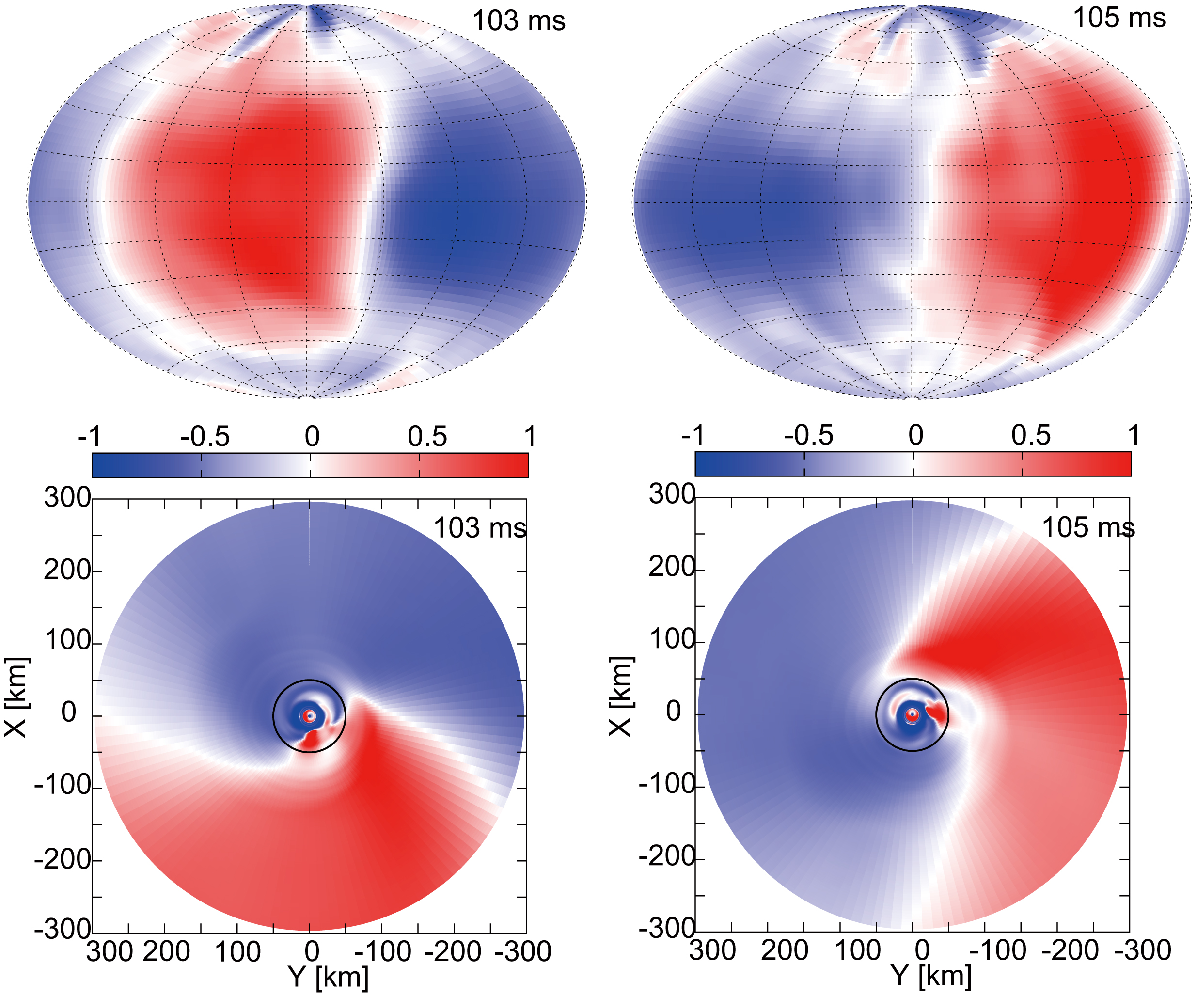}
\end{center}
 \caption{Rotation of the anisotropic $\bar{\nu}_e$ flux 
 around the spin axis. Top panels show the deviation of the flux 
 from the angle-averaged one ($\delta F_{\bar{\nu}_e}$, Eq.(\ref{eq1})) 
 at two snapshots of 103 ms (left panels) and 105 ms 
(right panels) after bounce, respectively. In the skymaps using
 the Hammer projection, the latitudes and longitudes are 
indicated by the dotted lines.
  Similar to the top panels, but the bottom panels display 
$\delta F_{\bar{\nu}_e}$ on the equatorial plane ($z=0$).
  The high neutrino emission region (colored by red) 
rotates in a counterclockwise direction
 from the bottom left to the bottom right panel (compare also the 
 top left with the top right panel).}\label{f2}
 \end{figure}

The clear signal modulation of Figure \ref{f1} originates from 
 a strong neutrino emitter rotating around the spin axis.
 This is essentially in analogue to a lighthouse effect as we 
 will explain in detail below.
 Figure \ref{f2} visualizes this, where
 the top two panels show how the excess in the anisotropic
  $\bar{\nu}_e$ flux (red regions) moves in the skymaps.
Here we estimate the degree of local anisotropic $\bar{\nu}_e$ flux as
\begin{equation}
  \delta F_{\bar{\nu}_e} =
   \frac{F_{\bar{\nu}_e} - \langle F_{\bar{\nu}_e} \rangle }
   {\sqrt{ \langle 
   \left( F_{\bar{\nu}_e}-\langle F_{\bar{\nu}_e} \rangle \right)^2
   }\rangle },
\label{eq1}
\end{equation}
where $F_{\bar{\nu}_e}$ is the flux measured at a radius of 500 km 
and $\langle F_{\bar{\nu}_e} \rangle$ is the angle-averaged one.
 In the skymaps, the direction where the flux is strong (weak) 
is colored by red (blue).
 The left and right panels correspond to the snapshot at 
103 ms and 105 ms after bounce, respectively.  
 At 103 ms (top left panel), the high neutrino
 emission region (colored by red) looks mostly concentrated 
in the western side in the skymap. More precisely, the highest 
neutrino emission comes from the direction near at the center
of the skymap $(\theta, \phi) \approx (\frac{\pi}{2}, \pi)$. And the 
high emission region is shown to move to the right (or eastern) 
direction in the top right panel at 105 ms after bounce,
 where the highest emission region is shifting to the direction at
$(\theta, \phi) \approx (\frac{\pi}{2}, \frac{3\pi}{2})$ due to 
rotation of the hot spot.
 The neutrino event rates become biggest
when the hot spot directly points to the observer.

 Looking into more details, we show that the time modulation of Figure \ref{f1} 
originally comes from rotation of the anisotropic $\bar{\nu}_e$ 
flux at the neutrino sphere.  The bottom panels of Figure \ref{f2} show
 the degree of the anisotropic $\bar{\nu}_e$ flux ($\delta F_{\bar{\nu}_e}$)
 in the equatorial plane (i.e., seen from 
the direction parallel to the spin axis). 
In the panels, the black circle drawn at a radius of 50 km 
roughly corresponds to the (average) $\bar{\nu}_e$ sphere. 
 In the bottom left panel,
 the excess of $\delta F_{\bar{\nu}_e}$ (colored by red) is strongest at  $(X,Y)=(-300,0)$, 
 corresponding to $(\theta, \phi) \approx (\frac{\pi}{2}, \pi)$ in 
the top left panel. At 2 ms later, the bottom right panel shows
 that the central region rotates counterclockwisely, making the 
hot spot rotate simultaneously to the right direction 
at $(X,Y)=(0,-300)$. Again
 this corresponds to the red region in the top right panel at 
$(\theta, \phi) \approx (\frac{\pi}{2}, \frac{3\pi}{2})$.
 
 The rotation of the anisotropic neutrino emission seen in Figure \ref{f2} 
is associated with the one-armed spiral flows. 
 As already discussed in \citet{takiwaki16}, the 
growth of the spiral flows 
 is triggered by the low-$T/|W|$ instability.
Like a screw, the non-axisymmetric flows initially develop 
from the PNS surface at a radius of 20 km (e.g., middle panel of Figure 3 
of \cite{takiwaki16}), then penetrate into the 
 $\bar{\nu}_e$ sphere 
(e.g., black circle in the bottom panels of Figure \ref{f2}),
 leading to the anisotropic neutrino emission.
 Rotation of the anisotropic $\bar{\nu}_e$ flux at the 
neutrino sphere is the origin of the {\it neutrino} lighthouse effect, leading
 to the observable modulation in the neutrino signal as shown in 
 Figure \ref{f1}.

\begin{figure}
\begin{center}
\includegraphics[width=0.99\linewidth]{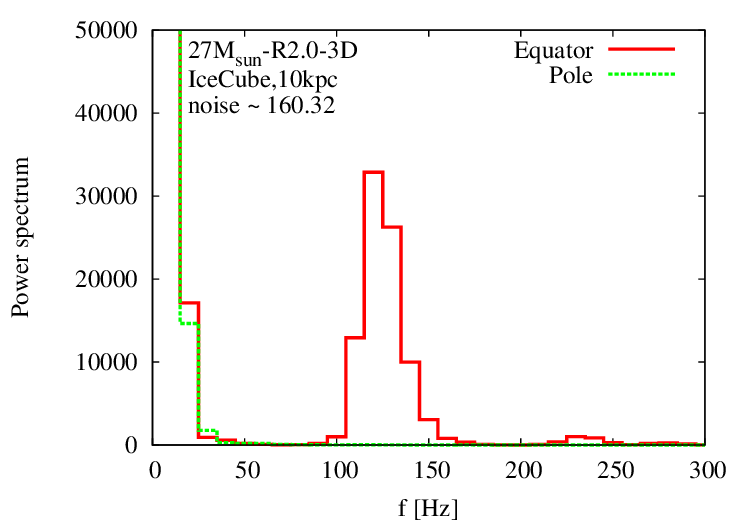}
\includegraphics[width=0.99\linewidth]{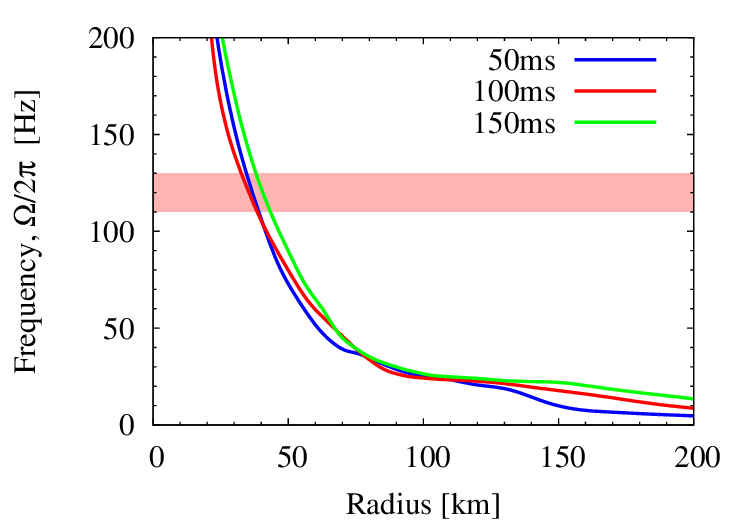}\\
\end{center}
 \caption{Similar to Figure \ref{f1} ({\it lower}) but for the
 power spectrum of the IceCube event rate 
 on the interval of 100 ms (from 50 to 150 ms after bounce).
 A pronounced peak can be seen at $\sim$ 120 Hz (red line) 
for an observer in the equatorial direction (labeled as "Equator"
). Bottom panel shows radial profiles of the 
rotational frequency of the matter at three postbounce times 
of 50 ms (blue line), 100 ms (red line) and 150 ms (green line), respectively.
The rotational frequency is estimated from the (azimuthally-averaged) 
angular velocity as $\Omega/2\pi$. 
The horizontal pink band approximately corresponds to the peak
 frequency in the spectrum with the HWHM contribution ($120 \pm 10$ Hz).}\label{f3}
\end{figure}

The top panel of Figure \ref{f3} shows
power spectrum of the IceCube event rate on the interval of 100 ms after bounce.
 Seen from the equator (red line), one can clearly see a peak around 120 Hz,
 which cannot be seen from the pole (green line).
Following \citet{lund10} (e.g., their Appendix A and Eq. (12)),
 the noise level of the IceCube is estimated as 160.32
  (for the 100 ms time interval). This is significantly lower
 than the peak amplitude (red line) at a 10 kpc distance scale.
 
 The bottom panel of Figure \ref{f3} shows that 
 the matter rotational frequency (at three different postbounce
 times) closely matches with the neutrino modulation frequency 
(horizontal pink band, $120 \pm 10$ Hz) at a radius
  of $40 \sim 50 $ km, which corresponds to 
the $\bar{\nu}_e$ sphere radius. 
This also supports the validity of the concept of the 
{\it neutrino} lighthouse effect firstly proposed in this work.
 

\begin{figure}
\begin{center}
\includegraphics[width=0.99\linewidth]{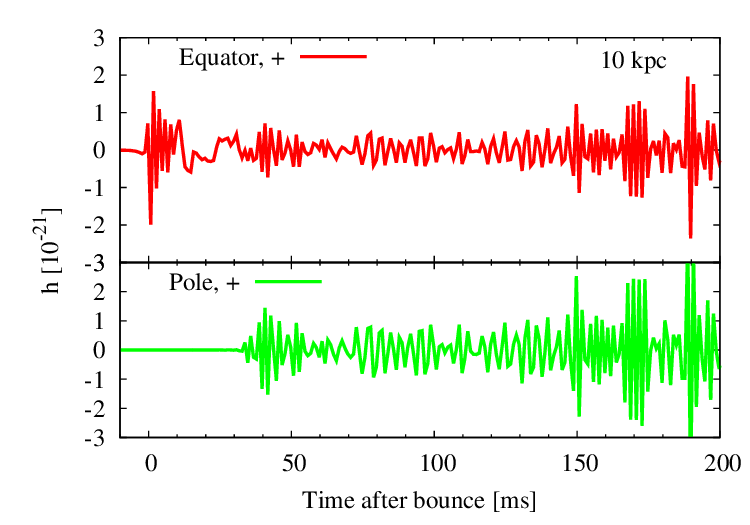}
\includegraphics[width=0.99\linewidth]{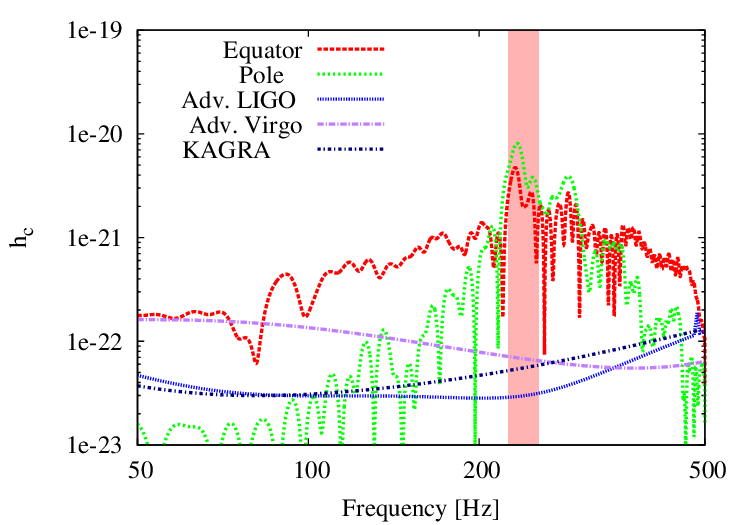}\\
\end{center}
 \caption{Top panel shows the GW amplitude of plus mode 
for observers along the equator (red line) and the pole (green line)
 for a source at 10 kpc. 
The bottom panel shows the 
characteristic GW spectra 
 relative to the sensitivity curves of advanced LIGO \citep{Harry10},
 advanced Virgo \citep{advv} 
and KAGRA \citep{Aso13}. Note in the bottom plot, 
we focus on only the frequency range ($\lesssim$ 500 Hz)
 with the maximum sensitivity.}\label{f4}
\end{figure}

Finally we shortly discuss the GW signatures from our 3D model. 
The top panel of Figure \ref{f4} shows the GW amplitude (plus mode) 
for observers along the equator (red line) and along the pole 
(green line), respectively. Seen from the equator (red line), 
the waveform after bounce is characterized by a big spike 
followed by the ring-down phase (up to $\sim$ 30 ms after bounce).
 This is the most generic waveform from rapidly rotating 
core-collapse and bounce \citep{Dimmelmeier08}, 
also known as the type I waveform. Seen from the spin
 axis (green line), the GW amplitude deviates from 
zero only after $\sim 40$ ms postbounce when the 
non-axisymmetric instabilities start to develop
 most strongly in the equatorial region. 
 The wave amplitude is thus bigger for the polar 
observer compared to the equatorial observer. 
Note that the polar-to-equator contrast in the GW amplitude is 
 stronger for the cross mode (not shown in the plot)
 than the plus mode. These features are consistent 
with those obtained in previous 
3D models \citep{Scheidegger10,Ott09} but with 
more approximate neutrino treatments.

The bottom panel of Figure \ref{f4} shows 
 the characteristic GW spectra ($h_c$) relative to 
the sensitivity curves of advanced LIGO and KAGRA for a source 
 at 10 kpc. Seen either from the equator 
(red line) or the pole (green line), the peak GW frequency 
is around 240Hz (e.g., the pink vertical band). 
Note that this is about twice as large as that of
 the modulation frequency of the neutrino signal ($\sim$ 120 Hz).
 This is simply because a deformed object rotating 
around the spin axis in the frequency of $f$ 
emits the quadrupole GW radiation with the frequency of $2f$.
 Our results show that the non-axisymmetric flow induced 
by the low-T/|W| instability is the key to explain the correlation
 between the time modulation in the neutrino and GW signal
 in the context of rapidly rotating CCSNe. 
One can also see that the characteristic GW 
amplitude for the polar observer (green line) in absence of the 
bounce GW signal is more narrow-banded compared to that for the 
equatorial observer (red line). As one would expect, 
the maximum $h_c$ at the peak GW frequency ($\sim$ 240 Hz)
 is bigger for the polar 
observer than that for the equatorial observer. It should be 
 noted that the peak frequency is close to the maximum 
sensitivities of advanced LIGO, advanced Virgo, and KAGRA.
These neutrino and GW signatures,
 if simultaneously detected, could provide new evidence 
 of rapid rotation of the forming PNS.

\section{Discussion}\label{sec:summary}

For enhancing our predictive power of the neutrino and GW 
 signals, we have to update our 
 numerical schemes in many respects. These include 
general relativistic effects with an improved multipole 
approximation of gravity \citep{couch13g}, multi-angle treatment
 in neutrino transport 
(e.g., \citet{Sumiyoshi15})
 beyond our ray-by-ray approximation using more detailed 
 neutrino opacities (e.g., \citet{Burrows06}). Given the rapidly rotating PNS, 
 magnetic fields should be also taken into account 
(e.g., \citet{moesta15}), where the field 
amplification due to the magneto-rotational instability 
(e.g., \citet{ober09,masada15}) could potentially affect
 the explosion dynamics. To
test this, a high-resolution 3D MHD model is needed, which
is another major undertaking. Finally it is noted that 
in the 3D model of the 11.2 $M_{\sun}$ progenitor \citep{takiwaki16} 
 no clear signatures as in Figures \ref{f1} and \ref{f4} are obtained. 
This is because the 11.2 $M_{\sun}$ model explodes even 
 without rotation where the low T/|W| instability does not
 develop during the simulation time.
 In order to clarify how general the correlated signatures found in this work 
would be, systematic 3D simulations  changing 
 the progenitor masses and initial rotation rates are needed, which we leave 
as a future work.

 \vspace{-0.5cm}
 \section{Acknowledgements}
The computations in this research were performed 
 on the K computer of
 the RIKEN AICS through 
the HPCI System Research project (Project ID:hp120304) together with 
XC30 of CfCA in NAOJ.
This study was supported by JSPS KAKENHI Grant Number 
(JP15H01039,  JP15H00789, JP15KK0173, JP17H06364, 
JP17K14306, and JP17H01130) and JICFuS as a priority issue to 
 be tackled by using Post 'K' computer.
\vspace{-0.5cm}




\bibliographystyle{mnras}
\bibliography{mybib} 

\begin{thebibliography}{}
\makeatletter
\relax
\def\mn@urlcharsother{\let\do\@makeother \do\$\do\&\do\#\do\^\do\_\do\%\do\~}
\def\mn@doi{\begingroup\mn@urlcharsother \@ifnextchar [ {\mn@doi@}
  {\mn@doi@[]}}
\def\mn@doi@[#1]#2{\def\@tempa{#1}\ifx\@tempa\@empty \href
  {http://dx.doi.org/#2} {doi:#2}\else \href {http://dx.doi.org/#2} {#1}\fi
  \endgroup}
\def\mn@eprint#1#2{\mn@eprint@#1:#2::\@nil}
\def\mn@eprint@arXiv#1{\href {http://arxiv.org/abs/#1} {{\tt arXiv:#1}}}
\def\mn@eprint@dblp#1{\href {http://dblp.uni-trier.de/rec/bibtex/#1.xml}
  {dblp:#1}}
\def\mn@eprint@#1:#2:#3:#4\@nil{\def\@tempa {#1}\def\@tempb {#2}\def\@tempc
  {#3}\ifx \@tempc \@empty \let \@tempc \@tempb \let \@tempb \@tempa \fi \ifx
  \@tempb \@empty \def\@tempb {arXiv}\fi \@ifundefined
  {mn@eprint@\@tempb}{\@tempb:\@tempc}{\expandafter \expandafter \csname
  mn@eprint@\@tempb\endcsname \expandafter{\@tempc}}}

\bibitem[\protect\citeauthoryear{{Abbasi} et~al.,}{{Abbasi}
  et~al.}{2011}]{ICECUBE2011}
{Abbasi} R.,  et~al., 2011, \mn@doi [\aap] {10.1051/0004-6361/201117810}, \href
  {http://adsabs.harvard.edu/abs/2011A%26A...535A.109A} {535, A109}

\bibitem[\protect\citeauthoryear{{Abbott} et~al.,}{{Abbott}
  et~al.}{2016}]{gw2016}
{Abbott} B.~P.,  et~al., 2016, \mn@doi [Physical Review Letters]
  {10.1103/PhysRevLett.116.061102}, \href
  {http://ads.nao.ac.jp/abs/2016PhRvL.116f1102A} {116, 061102}

\bibitem[\protect\citeauthoryear{{Abe} et~al.,}{{Abe}
  et~al.}{2011}]{HyperK2011}
{Abe} K.,  et~al., 2011, preprint, \href
  {http://adsabs.harvard.edu/abs/2011arXiv1109.3262A} {} (\mn@eprint {arXiv}
  {1109.3262})

\bibitem[\protect\citeauthoryear{{Aso}, {Michimura}, {Somiya}, {Ando},
  {Miyakawa}, {Sekiguchi}, {Tatsumi}  \& {Yamamoto}}{{Aso}
  et~al.}{2013}]{Aso13}
{Aso} Y.,  {Michimura} Y.,  {Somiya} K.,  {Ando} M.,  {Miyakawa} O.,
  {Sekiguchi} T.,  {Tatsumi} D.,   {Yamamoto} H.,  2013, \mn@doi [\prd]
  {10.1103/PhysRevD.88.043007}, \href
  {http://cdsads.u-strasbg.fr/abs/2013PhRvD..88d3007A} {88, 043007}

\bibitem[\protect\citeauthoryear{{Bethe}}{{Bethe}}{1990}]{bethe}
{Bethe} H.~A.,  1990, \mn@doi [Reviews of Modern Physics]
  {10.1103/RevModPhys.62.801}, \href
  {http://ads.nao.ac.jp/abs/1990RvMP...62..801B} {62, 801}

\bibitem[\protect\citeauthoryear{{Bollig}, {Janka}, {Lohs}, {Martinez-Pinedo},
  {Horowitz}  \& {Melson}}{{Bollig} et~al.}{2017}]{bollig17}
{Bollig} R.,  {Janka} H.-T.,  {Lohs} A.,  {Martinez-Pinedo} G.,  {Horowitz}
  C.~J.,   {Melson} T.,  2017, preprint, \href
  {http://ads.nao.ac.jp/abs/2017arXiv170604630B} {} (\mn@eprint {arXiv}
  {1706.04630})

\bibitem[\protect\citeauthoryear{{Burrows}}{{Burrows}}{2013}]{burrows13}
{Burrows} A.,  2013, \mn@doi [Reviews of Modern Physics]
  {10.1103/RevModPhys.85.245}, \href
  {http://adsabs.harvard.edu/abs/2013RvMP...85..245B} {85, 245}

\bibitem[\protect\citeauthoryear{{Burrows}, {Reddy}  \& {Thompson}}{{Burrows}
  et~al.}{2006}]{Burrows06}
{Burrows} A.,  {Reddy} S.,   {Thompson} T.~A.,  2006, \mn@doi [Nuclear Physics
  A] {10.1016/j.nuclphysa.2004.06.012}, \href
  {http://cdsads.u-strasbg.fr/abs/2006NuPhA.777..356B} {777, 356}

\bibitem[\protect\citeauthoryear{{Couch} \& {Ott}}{{Couch} \&
  {Ott}}{2015}]{Couch15}
{Couch} S.~M.,  {Ott} C.~D.,  2015, \mn@doi [\apj] {10.1088/0004-637X/799/1/5},
  \href {http://ads.nao.ac.jp/abs/2015ApJ...799....5C} {799, 5}

\bibitem[\protect\citeauthoryear{{Couch}, {Graziani}  \& {Flocke}}{{Couch}
  et~al.}{2013}]{couch13g}
{Couch} S.~M.,  {Graziani} C.,   {Flocke} N.,  2013, \mn@doi [\apj]
  {10.1088/0004-637X/778/2/181}, \href
  {http://adsabs.harvard.edu/abs/2013ApJ...778..181C} {778, 181}

\bibitem[\protect\citeauthoryear{{Dimmelmeier}, {Ott}, {Marek}  \&
  {Janka}}{{Dimmelmeier} et~al.}{2008}]{Dimmelmeier08}
{Dimmelmeier} H.,  {Ott} C.~D.,  {Marek} A.,   {Janka} H.-T.,  2008, \mn@doi
  [\prd] {10.1103/PhysRevD.78.064056}, \href
  {http://ads.nao.ac.jp/abs/2008PhRvD..78f4056D} {78, 064056}

\bibitem[\protect\citeauthoryear{{Finn} \& {Evans}}{{Finn} \&
  {Evans}}{1990}]{finn90}
{Finn} L.~S.,  {Evans} C.~R.,  1990, \mn@doi [\apj] {10.1086/168497}, \href
  {http://ads.nao.ac.jp/abs/1990ApJ...351..588F} {351, 588}

\bibitem[\protect\citeauthoryear{Fukugita \& Yanagida}{Fukugita \&
  Yanagida}{2003}]{fukugita03}
Fukugita M.,  Yanagida T.,  2003, Physics of Neutrino, 2003 edn.
Springer

\bibitem[\protect\citeauthoryear{{Guilet} \& {M{\"u}ller}}{{Guilet} \&
  {M{\"u}ller}}{2015}]{jerome15}
{Guilet} J.,  {M{\"u}ller} E.,  2015, \mn@doi [\mnras] {10.1093/mnras/stv727},
  \href {http://ads.nao.ac.jp/abs/2015MNRAS.450.2153G} {450, 2153}

\bibitem[\protect\citeauthoryear{{Hanke}, {M{\"u}ller}, {Wongwathanarat},
  {Marek}  \& {Janka}}{{Hanke} et~al.}{2013}]{Hanke13}
{Hanke} F.,  {M{\"u}ller} B.,  {Wongwathanarat} A.,  {Marek} A.,   {Janka}
  H.-T.,  2013, \mn@doi [\apj] {10.1088/0004-637X/770/1/66}, \href
  {http://cdsads.u-strasbg.fr/abs/2013ApJ...770...66H} {770, 66}

\bibitem[\protect\citeauthoryear{{Harry} \& {LIGO Scientific
  Collaboration}}{{Harry} \& {LIGO Scientific Collaboration}}{2010}]{Harry10}
{Harry} G.~M.,  {LIGO Scientific Collaboration} 2010, \mn@doi [Classical and
  Quantum Gravity] {10.1088/0264-9381/27/8/084006}, \href
  {http://ads.nao.ac.jp/abs/2010CQGra..27h4006H} {27, 084006}

\bibitem[\protect\citeauthoryear{{Hild}, {Freise}, {Mantovani}, {Chelkowski},
  {Degallaix}  \& {Schilling}}{{Hild} et~al.}{2009}]{advv}
{Hild} S.,  {Freise} A.,  {Mantovani} M.,  {Chelkowski} S.,  {Degallaix} J.,
  {Schilling} R.,  2009, \mn@doi [Classical and Quantum Gravity]
  {10.1088/0264-9381/26/2/025005}, \href
  {http://ads.nao.ac.jp/abs/2009CQGra..26b5005H} {26, 025005}

\bibitem[\protect\citeauthoryear{{Hyper-Kamiokande
  proto-collaboration}}{{Hyper-Kamiokande
  proto-collaboration}}{2016}]{HyperK2016}
{Hyper-Kamiokande proto-collaboration} 2016, KEK-PREPRINT-2016-21,
  ICRR-REPORT-701-2016-1

\bibitem[\protect\citeauthoryear{{Janka}, {Melson}  \& {Summa}}{{Janka}
  et~al.}{2016}]{janka16}
{Janka} H.-T.,  {Melson} T.,   {Summa} A.,  2016, Annual Review of Nuclear and
  Particle Science, \href {http://adsabs.harvard.edu/abs/2016arXiv160205576J}
  {}

\bibitem[\protect\citeauthoryear{{Keil}, {Raffelt}  \& {Janka}}{{Keil}
  et~al.}{2003}]{keil03}
{Keil} M.~T.,  {Raffelt} G.~G.,   {Janka} H.-T.,  2003, \mn@doi [\apj]
  {10.1086/375130}, \href {http://adsabs.harvard.edu/abs/2003ApJ...590..971K}
  {590, 971}

\bibitem[\protect\citeauthoryear{{Kotake}}{{Kotake}}{2013}]{Kotake13}
{Kotake} K.,  2013, \mn@doi [Comptes Rendus Physique]
  {10.1016/j.crhy.2013.01.008}, \href
  {http://adsabs.harvard.edu/abs/2013CRPhy..14..318K} {14, 318}

\bibitem[\protect\citeauthoryear{{Kuroda}, {Kotake}  \& {Takiwaki}}{{Kuroda}
  et~al.}{2012}]{KurodaT12}
{Kuroda} T.,  {Kotake} K.,   {Takiwaki} T.,  2012, \mn@doi [\apj]
  {10.1088/0004-637X/755/1/11}, \href
  {http://ads.nao.ac.jp/abs/2012ApJ...755...11K} {755, 11}

\bibitem[\protect\citeauthoryear{{Kuroda}, {Takiwaki}  \& {Kotake}}{{Kuroda}
  et~al.}{2014}]{KurodaT14}
{Kuroda} T.,  {Takiwaki} T.,   {Kotake} K.,  2014, \mn@doi [\prd]
  {10.1103/PhysRevD.89.044011}, \href
  {http://ads.nao.ac.jp/abs/2014PhRvD..89d4011K} {89, 044011}

\bibitem[\protect\citeauthoryear{{Lentz} et~al.,}{{Lentz}
  et~al.}{2015}]{lentz15}
{Lentz} E.~J.,  et~al., 2015, \mn@doi [\apjl] {10.1088/2041-8205/807/2/L31},
  \href {http://ads.nao.ac.jp/abs/2015ApJ...807L..31L} {807, L31}

\bibitem[\protect\citeauthoryear{{Liebend{\"o}rfer}, {Whitehouse}  \&
  {Fischer}}{{Liebend{\"o}rfer} et~al.}{2009}]{idsa}
{Liebend{\"o}rfer} M.,  {Whitehouse} S.~C.,   {Fischer} T.,  2009, \mn@doi
  [\apj] {10.1088/0004-637X/698/2/1174}, \href
  {http://ads.nao.ac.jp/abs/2009ApJ...698.1174L} {698, 1174}

\bibitem[\protect\citeauthoryear{{Lund}, {Marek}, {Lunardini}, {Janka}  \&
  {Raffelt}}{{Lund} et~al.}{2010}]{lund10}
{Lund} T.,  {Marek} A.,  {Lunardini} C.,  {Janka} H.-T.,   {Raffelt} G.,  2010,
  \mn@doi [\prd] {10.1103/PhysRevD.82.063007}, \href
  {http://adsabs.harvard.edu/abs/2010PhRvD..82f3007L} {82, 063007}

\bibitem[\protect\citeauthoryear{{Marek} \& {Janka}}{{Marek} \&
  {Janka}}{2009}]{Marek09}
{Marek} A.,  {Janka} H.-T.,  2009, \mn@doi [\apj]
  {10.1088/0004-637X/694/1/664}, \href
  {http://ads.nao.ac.jp/abs/2009ApJ...694..664M} {694, 664}

\bibitem[\protect\citeauthoryear{{Masada}, {Takiwaki}  \& {Kotake}}{{Masada}
  et~al.}{2015}]{masada15}
{Masada} Y.,  {Takiwaki} T.,   {Kotake} K.,  2015, \mn@doi [\apjl]
  {10.1088/2041-8205/798/1/L22}, \href
  {http://adsabs.harvard.edu/abs/2015ApJ...798L..22M} {798, L22}

\bibitem[\protect\citeauthoryear{{Melson}, {Janka}, {Bollig}, {Hanke}, {Marek}
  \& {M{\"u}ller}}{{Melson} et~al.}{2015}]{melson15b}
{Melson} T.,  {Janka} H.-T.,  {Bollig} R.,  {Hanke} F.,  {Marek} A.,
  {M{\"u}ller} B.,  2015, \mn@doi [\apjl] {10.1088/2041-8205/808/2/L42}, \href
  {http://ads.nao.ac.jp/abs/2015ApJ...808L..42M} {808, L42}

\bibitem[\protect\citeauthoryear{{Mirizzi}, {Tamborra}, {Janka}, {Saviano},
  {Scholberg}, {Bollig}, {H{\"u}depohl}  \& {Chakraborty}}{{Mirizzi}
  et~al.}{2016}]{mirizzi16}
{Mirizzi} A.,  {Tamborra} I.,  {Janka} H.-T.,  {Saviano} N.,  {Scholberg} K.,
  {Bollig} R.,  {H{\"u}depohl} L.,   {Chakraborty} S.,  2016, \mn@doi [Nuovo
  Cimento Rivista Serie] {10.1393/ncr/i2016-10120-8}, \href
  {http://ads.nao.ac.jp/abs/2016NCimR..39....1M} {39, 1}

\bibitem[\protect\citeauthoryear{{Misner}, {Thorne}  \& {Wheeler}}{{Misner}
  et~al.}{1973}]{Misner73}
{Misner} C.~W.,  {Thorne} K.~S.,   {Wheeler} J.~A.,  1973, {Gravitation}.
Princeton University Press

\bibitem[\protect\citeauthoryear{{M{\"o}sta}, {Ott}, {Radice}, {Roberts},
  {Schnetter}  \& {Haas}}{{M{\"o}sta} et~al.}{2015}]{moesta15}
{M{\"o}sta} P.,  {Ott} C.~D.,  {Radice} D.,  {Roberts} L.~F.,  {Schnetter} E.,
   {Haas} R.,  2015, \mn@doi [\nat] {10.1038/nature15755}, \href
  {http://adsabs.harvard.edu/abs/2015Natur.528..376M} {528, 376}

\bibitem[\protect\citeauthoryear{{M\"uller} \& {Janka}}{{M\"uller} \&
  {Janka}}{1997}]{EMuller97}
{M\"uller} E.,  {Janka} H.-T.,  1997, \aap, \href
  {http://ads.nao.ac.jp/abs/1997A%26A...317..140M} {317, 140}

\bibitem[\protect\citeauthoryear{{M{\"u}ller} \& {Janka}}{{M{\"u}ller} \&
  {Janka}}{2015}]{bernhard15i}
{M{\"u}ller} B.,  {Janka} H.-T.,  2015, \mn@doi [\mnras]
  {10.1093/mnras/stv101}, \href {http://ads.nao.ac.jp/abs/2015MNRAS.448.2141M}
  {448, 2141}

\bibitem[\protect\citeauthoryear{{M{\"u}ller}, {Janka}  \&
  {Heger}}{{M{\"u}ller} et~al.}{2012}]{BMuller12b}
{M{\"u}ller} B.,  {Janka} H.-T.,   {Heger} A.,  2012, \mn@doi [\apj]
  {10.1088/0004-637X/761/1/72}, \href
  {http://ads.nao.ac.jp/abs/2012ApJ...761...72M} {761, 72}

\bibitem[\protect\citeauthoryear{{M{\"u}ller}, {Janka}  \&
  {Marek}}{{M{\"u}ller} et~al.}{2013}]{BMuller13}
{M{\"u}ller} B.,  {Janka} H.-T.,   {Marek} A.,  2013, \mn@doi [\apj]
  {10.1088/0004-637X/766/1/43}, \href
  {http://ads.nao.ac.jp/abs/2013ApJ...766...43M} {766, 43}

\bibitem[\protect\citeauthoryear{{M{\"u}ller}, {Melson}, {Heger}  \&
  {Janka}}{{M{\"u}ller} et~al.}{2017}]{bernhard17}
{M{\"u}ller} B.,  {Melson} T.,  {Heger} A.,   {Janka} H.-T.,  2017, \mn@doi
  [\mnras] {10.1093/mnras/stx1962}, \href
  {http://ads.nao.ac.jp/abs/2017MNRAS.472..491M} {472, 491}

\bibitem[\protect\citeauthoryear{{Murphy}, {Ott}  \& {Burrows}}{{Murphy}
  et~al.}{2009}]{Murphy09}
{Murphy} J.~W.,  {Ott} C.~D.,   {Burrows} A.,  2009, \mn@doi [\apj]
  {10.1088/0004-637X/707/2/1173}, \href
  {http://ads.nao.ac.jp/abs/2009ApJ...707.1173M} {707, 1173}

\bibitem[\protect\citeauthoryear{{Nakamura}, {Takiwaki}, {Kuroda}  \&
  {Kotake}}{{Nakamura} et~al.}{2015}]{Nakamura15}
{Nakamura} K.,  {Takiwaki} T.,  {Kuroda} T.,   {Kotake} K.,  2015, \mn@doi
  [\pasj] {10.1093/pasj/psv073}, \href
  {http://cdsads.u-strasbg.fr/abs/2015PASJ...67..107N} {67, 107}

\bibitem[\protect\citeauthoryear{{Obergaulinger} \& {Aloy}}{{Obergaulinger} \&
  {Aloy}}{2017}]{martin17}
{Obergaulinger} M.,  {Aloy} M.~{\'A}.,  2017, \mn@doi [\mnras]
  {10.1093/mnrasl/slx046}, \href {http://ads.nao.ac.jp/abs/2017MNRAS.469L..43O}
  {469, L43}

\bibitem[\protect\citeauthoryear{{Obergaulinger}, {Cerd{\'a}-Dur{\'a}n},
  {M{\"u}ller}  \& {Aloy}}{{Obergaulinger} et~al.}{2009}]{ober09}
{Obergaulinger} M.,  {Cerd{\'a}-Dur{\'a}n} P.,  {M{\"u}ller} E.,   {Aloy}
  M.~A.,  2009, \mn@doi [\aap] {10.1051/0004-6361/200811323}, \href
  {http://ads.nao.ac.jp/abs/2009A%26A...498..241O} {498, 241}

\bibitem[\protect\citeauthoryear{{Ott}}{{Ott}}{2009}]{Ott09}
{Ott} C.~D.,  2009, \mn@doi [Classical and Quantum Gravity]
  {10.1088/0264-9381/26/6/063001}, \href
  {http://ads.nao.ac.jp/abs/2009CQGra..26f3001O} {26, 063001}

\bibitem[\protect\citeauthoryear{{Ott} et~al.,}{{Ott} et~al.}{2012}]{Ott12a}
{Ott} C.~D.,  et~al., 2012, \mn@doi [\prd] {10.1103/PhysRevD.86.024026}, \href
  {http://ads.nao.ac.jp/abs/2012PhRvD..86b4026O} {86, 024026}

\bibitem[\protect\citeauthoryear{{Rembiasz}, {Guilet}, {Obergaulinger},
  {Cerd{\'a}-Dur{\'a}n}, {Aloy}  \& {M{\"u}ller}}{{Rembiasz}
  et~al.}{2016}]{tomek16}
{Rembiasz} T.,  {Guilet} J.,  {Obergaulinger} M.,  {Cerd{\'a}-Dur{\'a}n} P.,
  {Aloy} M.~A.,   {M{\"u}ller} E.,  2016, \mn@doi [\mnras]
  {10.1093/mnras/stw1201}, \href {http://ads.nao.ac.jp/abs/2016MNRAS.460.3316R}
  {460, 3316}

\bibitem[\protect\citeauthoryear{{Roberts}, {Ott}, {Haas}, {O'Connor}, {Diener}
   \& {Schnetter}}{{Roberts} et~al.}{2016}]{Roberts16}
{Roberts} L.~F.,  {Ott} C.~D.,  {Haas} R.,  {O'Connor} E.~P.,  {Diener} P.,
  {Schnetter} E.,  2016, \mn@doi [\apj] {10.3847/0004-637X/831/1/98}, \href
  {http://ads.nao.ac.jp/abs/2016ApJ...831...98R} {831, 98}

\bibitem[\protect\citeauthoryear{{Salathe}, {Ribordy}  \&
  {Demir{\"o}rs}}{{Salathe} et~al.}{2012}]{ICECUBE2012}
{Salathe} M.,  {Ribordy} M.,   {Demir{\"o}rs} L.,  2012, \mn@doi [Astroparticle
  Physics] {10.1016/j.astropartphys.2011.10.012}, \href
  {http://adsabs.harvard.edu/abs/2012APh....35..485S} {35, 485}

\bibitem[\protect\citeauthoryear{{Scheidegger}, {K{\"a}ppeli}, {Whitehouse},
  {Fischer}  \& {Liebend{\"o}rfer}}{{Scheidegger} et~al.}{2010}]{Scheidegger10}
{Scheidegger} S.,  {K{\"a}ppeli} R.,  {Whitehouse} S.~C.,  {Fischer} T.,
  {Liebend{\"o}rfer} M.,  2010, \mn@doi [\aap] {10.1051/0004-6361/200913220},
  \href {http://cdsads.u-strasbg.fr/abs/2010A%26A...514A..51S} {514, A51}

\bibitem[\protect\citeauthoryear{{Scholberg}}{{Scholberg}}{2012}]{scholberg12}
{Scholberg} K.,  2012, \mn@doi [Annual Review of Nuclear and Particle Science]
  {10.1146/annurev-nucl-102711-095006}, \href
  {http://adsabs.harvard.edu/abs/2012ARNPS..62...81S} {62, 81}

\bibitem[\protect\citeauthoryear{{Sumiyoshi}, {Takiwaki}, {Matsufuru}  \&
  {Yamada}}{{Sumiyoshi} et~al.}{2015}]{Sumiyoshi15}
{Sumiyoshi} K.,  {Takiwaki} T.,  {Matsufuru} H.,   {Yamada} S.,  2015, \mn@doi
  [\apjs] {10.1088/0067-0049/216/1/5}, \href
  {http://ads.nao.ac.jp/abs/2015ApJS..216....5S} {216, 5}

\bibitem[\protect\citeauthoryear{{Summa}, {Janka}, {Melson}  \&
  {Marek}}{{Summa} et~al.}{2017}]{summa17}
{Summa} A.,  {Janka} H.-T.,  {Melson} T.,   {Marek} A.,  2017, preprint, \href
  {http://ads.nao.ac.jp/abs/2017arXiv170804154S} {} (\mn@eprint {arXiv}
  {1708.04154})

\bibitem[\protect\citeauthoryear{{Suwa}, {Kotake}, {Takiwaki}, {Whitehouse},
  {Liebend{\"o}rfer}  \& {Sato}}{{Suwa} et~al.}{2010}]{Suwa10}
{Suwa} Y.,  {Kotake} K.,  {Takiwaki} T.,  {Whitehouse} S.~C.,
  {Liebend{\"o}rfer} M.,   {Sato} K.,  2010, \mn@doi [\pasj]
  {10.1093/pasj/62.6.L49}, \href {http://ads.nao.ac.jp/abs/2010PASJ...62L..49S}
  {62, L49}

\bibitem[\protect\citeauthoryear{{Takiwaki} \& {Kotake}}{{Takiwaki} \&
  {Kotake}}{2011}]{tk11}
{Takiwaki} T.,  {Kotake} K.,  2011, \mn@doi [\apj]
  {10.1088/0004-637X/743/1/30}, \href
  {http://adsabs.harvard.edu/abs/2011ApJ...743...30T} {743, 30}

\bibitem[\protect\citeauthoryear{{Takiwaki}, {Kotake}  \& {Suwa}}{{Takiwaki}
  et~al.}{2016}]{takiwaki16}
{Takiwaki} T.,  {Kotake} K.,   {Suwa} Y.,  2016, \mn@doi [\mnras]
  {10.1093/mnrasl/slw105}, \href
  {http://adsabs.harvard.edu/abs/2016MNRAS.461L.112T} {461, L112}

\bibitem[\protect\citeauthoryear{{Tamborra}, {Hanke}, {M{\"u}ller}, {Janka}  \&
  {Raffelt}}{{Tamborra} et~al.}{2013}]{tamborra13}
{Tamborra} I.,  {Hanke} F.,  {M{\"u}ller} B.,  {Janka} H.-T.,   {Raffelt} G.,
  2013, \mn@doi [Physical Review Letters] {10.1103/PhysRevLett.111.121104},
  \href {http://adsabs.harvard.edu/abs/2013PhRvL.111l1104T} {111, 121104}

\bibitem[\protect\citeauthoryear{{Tamborra}, {Raffelt}, {Hanke}, {Janka}  \&
  {M{\"u}ller}}{{Tamborra} et~al.}{2014}]{tamborra14a}
{Tamborra} I.,  {Raffelt} G.,  {Hanke} F.,  {Janka} H.-T.,   {M{\"u}ller} B.,
  2014, \mn@doi [\prd] {10.1103/PhysRevD.90.045032}, \href
  {http://adsabs.harvard.edu/abs/2014PhRvD..90d5032T} {90, 045032}

\bibitem[\protect\citeauthoryear{{The LIGO Scientific Collaboration and the
  Virgo Collaboration} et~al.,}{{The LIGO Scientific Collaboration and the
  Virgo Collaboration} et~al.}{2017}]{joint}
{The LIGO Scientific Collaboration and the Virgo Collaboration} et~al., 2017,
  preprint, \href {http://ads.nao.ac.jp/abs/2017arXiv170909660T} {} (\mn@eprint
  {arXiv} {1709.09660})

\bibitem[\protect\citeauthoryear{{Woosley}, {Heger}  \& {Weaver}}{{Woosley}
  et~al.}{2002}]{woosley02}
{Woosley} S.~E.,  {Heger} A.,   {Weaver} T.~A.,  2002, \mn@doi [Reviews of
  Modern Physics] {10.1103/RevModPhys.74.1015}, \href
  {http://ads.nao.ac.jp/abs/2002RvMP...74.1015W} {74, 1015}

\bibitem[\protect\citeauthoryear{{Yokozawa}, {Asano}, {Kayano}, {Suwa},
  {Kanda}, {Koshio}  \& {Vagins}}{{Yokozawa} et~al.}{2015}]{Yokozawa}
{Yokozawa} T.,  {Asano} M.,  {Kayano} T.,  {Suwa} Y.,  {Kanda} N.,  {Koshio}
  Y.,   {Vagins} M.~R.,  2015, \mn@doi [\apj] {10.1088/0004-637X/811/2/86},
  \href {http://ads.nao.ac.jp/abs/2015ApJ...811...86Y} {811, 86}

\makeatother
\end{thebibliography}


\label{lastpage}
\end{document}